\documentclass[reprint,amsmath,amssymb,pra,aps,superscriptaddress,nolongbibliography]{revtex4-2}
\usepackage[inline]{enumitem}
\usepackage{graphicx} 
\usepackage{tikz}
\usetikzlibrary{calc}

\begin{document}
\title{Matrix Product Evolution: A Method for Simulating Quantum Circuits Using Tensor Networks}
\author{Haruyuki Kawabe}
\affiliation{Technology Research and Innovation, BIPROGY Inc., Tokyo 135-8560, Japan}

\author{Minoru Nagai}
\affiliation{Technology Research and Innovation, BIPROGY Inc., Tokyo 135-8560, Japan}

\author{Tsuyoshi Okubo}
\affiliation{Department of Physics, University of Tokyo, Tokyo 113-0033, Japan}
\affiliation{Department of Physics, Niigata University, Niigata 950-2181, Japan}

\author{Synge Todo}
\affiliation{Department of Physics, University of Tokyo, Tokyo 113-0033, Japan}
\affiliation{Institute for Physics of Intelligence, University of Tokyo, Tokyo 113-0033, Japan}
\affiliation{Institute for Solid State Physics, University of Tokyo, Kashiwa 277-8581, Japan}

\begin{abstract}
Classical simulation of quantum circuits is an essential tool in quantum information science, but its applicability is constrained by the exponential growth of the Hilbert space and the entanglement structure of quantum states. 
In this work, we introduce Matrix Product Evolution (MPE), a tensor-train representation of quantum circuits constructed along the circuit depth rather than along the qubit index. 
Within this formulation, the simulation of a quantum circuit is modeled as the contraction of multiple MPE tensors. 
We develop an efficient contraction strategy based on a zip-up procedure to carry out this contraction in practice. 
We investigate the numerical behavior of this MPE-based contraction framework through simulations of random quantum circuits and the time evolution of a quantum many-body state.
Our results characterize the growth of temporal bond dimensions, clarify how post-selection modifies the contraction cost and approximation accuracy, and identify regimes in which depth-oriented tensor-network contractions provide a useful complement to standard MPS-based simulation approaches.
\end{abstract}

\maketitle

\section{Introduction}
The classical simulation of quantum circuits plays a central role in quantum information science, providing information on quantum dynamics and enabling the quantitative validation of quantum algorithms and devices \cite{Feynman1982, Lloyd1996, Nielsen2000, Preskill2018, Harrow2017}. 
Despite rapid progress in quantum hardware, classical simulation remains indispensable for understanding the structure of quantum states and for exploring regimes that are not yet accessible experimentally. However, the exponential growth of the Hilbert space with the number of qubits severely limits direct simulation methods based on full state vectors.

State-based tensor-network approaches such as TEBD exploit limited entanglement to simulate one-dimensional dynamics efficiently \cite{Vidal2003, Vidal2004, Vidal2007, OrusVidal2008}.
Tensor-network techniques offer a powerful framework for addressing this challenge by representing quantum states and operations in compressed forms that exploit their entanglement structure. 
Among these approaches, the matrix product state (MPS) representation has been widely used to simulate many-body quantum systems and quantum circuits with limited entanglement \cite{White1992, White1993, Ostlund1995, Rommer1997, Schollwock2005, Schollwock2011, Verstraete2008, PerezGarcia2007, Orus2014, Orus2019}.    
In the MPS-based framework, a many-qubit quantum state at a given time is represented as a one-dimensional tensor network along the spatial direction, and quantum gates are incorporated through sequential updates that propagate the time evolution of the state. 
This approach is effective when the entanglement across spatial bipartitions remains limited, allowing the bond dimensions to be truncated without a significant loss of accuracy.

A quantum circuit can be regarded as a two-dimensional tensor network, and different contraction orders lead to different computational trade-offs \cite{Zwolak2004, Paeckel2019, Haegeman2016, Markov2008, Shi2006, Bremner2011, Bravyi2016, Pednault2019}.
Standard MPS-based simulations correspond to a contraction strategy that prioritizes the explicit representation of the quantum state at each time step, effectively organizing correlations with respect to spatial bipartitions. 
Alternative contraction orders can therefore provide complementary viewpoints and computational advantages, depending on how correlations are distributed in the circuit.

In this work, we introduce Matrix Product Evolution (MPE) as a tensor representation of quantum circuits that reorganizes this contraction problem from the perspective of the temporal structure of the circuit. 
In contrast to MPS, which represents a quantum state at a fixed time, MPE represents the time evolution of a single qubit or a small set of neighboring qubits as a tensor train along the temporal direction. 
Based on this representation, we develop an MPE-based method in which the simulation of a quantum circuit is formulated as the successive contraction of multiple MPEs.

This formulation provides a complementary perspective to MPS-based approaches. 
While the bond dimension of an MPS reflects the Schmidt rank across a spatial bipartition, the temporal bond dimension of an MPE reflects correlations between earlier and later portions of the evolution history, which is closely related to notions of temporal entanglement studied in time-direction tensor-network representations and influence-matrix approaches~\cite{Lerose2021,Giudice2022,Foligno2023}.
As a result, its behavior depends on the temporal structure of the circuit rather than solely on
the amount of spatial entanglement present in the quantum state.

The computational behavior of the MPE-based method therefore differs from that of MPS-based simulations, particularly in how bond dimensions grow and how constraints such as post-selection affect the effective degrees of freedom.

We develop an efficient contraction scheme for this representation based on a zip-up procedure, which enables a practical evaluation of the tensor network associated with a quantum circuit.
Using this framework, we analyze the growth of bond dimensions along the temporal direction and clarify how post-selection modifies the bounds on these bond dimensions. 
We emphasize that this dependence reflects a structural feature of the MPE-based representation rather than a general advantage over MPS-based methods.

We then apply the MPE-based method to numerical simulations of random quantum circuits and quantum many-body time evolution, and compare its performance with that of the MPS-based approach. 
Our results elucidate the regimes in which the MPE-based method provides a useful complementary perspective, as well as its limitations when entanglement grows rapidly.


\section{METHODS}

\begin{figure*}[tbph]
\begin{tikzpicture}[T/.style={shape=rectangle, minimum width=0.35cm, minimum height=0.5cm, fill=white, draw},C/.style={shape=rectangle, minimum width=0.35cm, minimum height=1.2cm, fill=white, draw},inner sep=1pt,font=\scriptsize, xscale=0.6, yscale=0.7]
  \node at (-1,3.5) {(1)};
  \foreach \y in {0,1,2,3} \draw (0,\y) -- +(3.8,0);
  \node[shape=rectangle,fill=white,draw,rotate=-90,draw,minimum width=2.6cm,minimum height=0.4cm] at (0,1.5) {initial state};
  \node[T] at (1,0) {};
  \node[C] at (1,2.5) {};
  \node[C] at (2,0.5) {};
  \node[T] at (2,2) {};
  \node[C] at (3,1.5) {};

\begin{scope}[shift={(5.5,0)}]
  \node at (-1,3.5) {(2)};
  \foreach \y in {0,1,2,3} \draw (0,\y) -- +(3.8,0);
  \draw (0,0) -- (0,3);
  \draw (1,2) -- (1,3);
  \draw (2,0) -- (2,1);
  \draw (3,1) -- (3,2);
  \node[T] at (0,0) {};
  \node[T] at (0,1) {};
  \node[T] at (0,2) {};
  \node[T] at (0,3) {};
  \node[T] at (1,0) {};
  \node[T] at (1,1) {$I$};
  \node[T] at (1,2) {};
  \node[T] at (1,3) {};
  \node[T] at (2,0) {};
  \node[T] at (2,1) {};
  \node[T] at (2,2) {};
  \node[T] at (2,3) {$I$};
  \node[T] at (3,0) {$I$};
  \node[T] at (3,1) {};
  \node[T] at (3,2) {};
  \node[T] at (3,3) {$I$};

\begin{scope}[shift={(5.5,0)}]
  \node at (-1,3.5) {(3)};
  \foreach \y in {0,1,2,3} \draw (0,\y) -- +(2.8,0);
  \foreach \x in {0,1,2} \draw (\x,0) -- +(0,3);
  \node[T] at (0,0) {};
  \node[T] at (0,1) {};
  \node[T] at (0,2) {};
  \node[T] at (0,3) {};
  \node[T] at (1,0) {};
  \node[T] at (1,1) {};
  \node[T] at (1,2) {};
  \node[T] at (1,3) {};
  \node[T] at (2,0) {};
  \node[T] at (2,1) {};
  \node[T] at (2,2) {};
  \node[T] at (2,3) {};

\begin{scope}[shift={(4.5,0)}]
  \node at (-1,3.5) {(4)};
  \foreach \y in {0,1,2} \draw (0,\y) -- +(2.8,0);
  \foreach \x in {0,1,2} \draw (\x,0) -- +(0,2);
  \draw (2,3) -- +(0.8,0);
  \draw (2,2) -- +(0,1);
  \node[T] at (0,0) {};
  \node[T] at (0,1) {};
  \node[T] at (0,2) {};
  \node[T] at (1,0) {};
  \node[T] at (1,1) {};
  \node[T] at (1,2) {};
  \node[T] at (2,0) {};
  \node[T] at (2,1) {};
  \node[T] at (2,2) {};
  \node[T] at (2,3) {};

\begin{scope}[shift={(4.5,0)}]
  \node at (-1,3.5) {(5)};
  \foreach \y in {1,2,3} \draw (2,\y) -- +(0.8,0);
  \draw (0,0) -- +(2.8,0);
  \draw (2,0) -- +(0,3);
  \node[T] at (0,0) {};
  \node[T] at (1,0) {};
  \node[T] at (2,0) {};
  \node[T] at (2,1) {};
  \node[T] at (2,2) {};
  \node[T] at (2,3) {};

\begin{scope}[shift={(4.5,0)}]
  \node at (-1,3.5) {(6)};
  \foreach \y in {0,1,2,3} \draw (0,\y) -- +(0.8,0);
  \draw (0,0) -- +(0,3);
  \node[T] at (0,0) {};
  \node[T] at (0,1) {};
  \node[T] at (0,2) {};
  \node[T] at (0,3) {};
\end{scope}
\end{scope}
\end{scope}
\end{scope}
\end{scope}
\end{tikzpicture}
\caption[]{Overview of the tensor-network representation and contraction strategy underlying the MPE-based simulation for a quantum circuit.
(1) An input quantum circuit consisting of an initial state and a sequence of quantum gates. 
(2–6) Construction of a corresponding tensor network, insertion of identity tensors, temporal compression of the tensor network, successive contraction of MPEs, and final contraction yielding the output quantum state. }
\label{fig:overview}
\end{figure*}
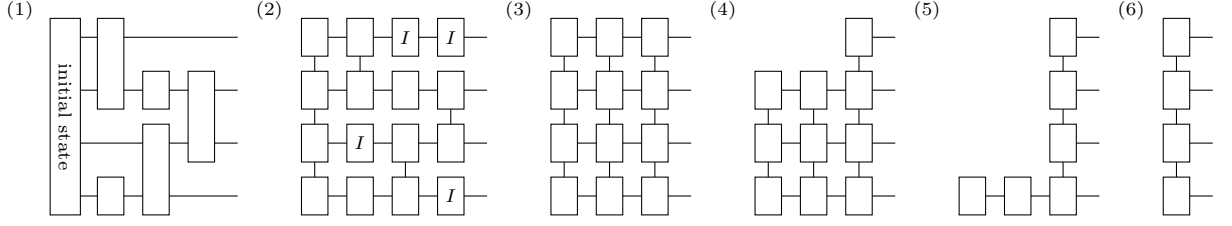

We consider quantum circuits acting on $n$ qubits. 
A circuit is represented as a two-dimensional tensor network whose spatial direction corresponds to the qubit index and whose temporal direction corresponds to the circuit depth. 
In conventional MPS-based simulations, this network is contracted by updating a quantum state represented along the spatial direction. 
In contrast, the method introduced here reorganizes the same tensor network along the temporal direction. 
The resulting representation, which we call Matrix Product Evolution (MPE), provides an approximation framework in which correlations accumulated during the circuit evolution are encoded in temporal bond dimensions. 

The overall procedure is summarized in Fig.~\ref{fig:overview}.
Starting from an input circuit and an initial quantum state, we first construct a rectangular two-dimensional tensor network by representing states and gates as local tensors and by inserting identity tensors where necessary. 
The network is then compressed along the temporal direction while preserving its rectangular structure. 
Finally, the resulting MPEs are contracted along the spatial direction by a zip-up procedure, yielding an MPS representation of the output state.

\subsection{Matrix Product Evolution}

Matrix Product Evolution (MPE) is a tensor-train representation organized along the temporal direction of a quantum circuit, analogously to matrix product states and tensor-train decompositions \cite{Verstraete2008, PerezGarcia2007, Orus2014, Orus2019, Oseledets2011, Fishman2022, Gray2018}.
Consider a subsystem $S$ consisting of one or more qubits. 
An MPE associated with $S$ is defined as a tensor train
\[ 
\mathcal{M(S)} = \{T_0, T_1, \ldots, T_d\}, 
\] 
where $d$ denotes the depth of the circuit and $T_k$ represents the local tensor corresponding to the $k$-th time step. 
The neighboring tensors are connected by internal bond indices along the temporal direction. 
These temporal bonds encode correlations accumulated during the circuit evolution and play a role analogous to the bond indices of a matrix product state (MPS).
 
In addition to temporal bond indices, each tensor may carry open indices originating from gates coupling $S$ to neighboring subsystems. These open indices connect different MPEs and are contracted during the spatial contraction procedure described in the following.

Initially, each row of the circuit tensor network, corresponding to the time evolution of a single qubit, forms an individual MPE. 
During the contraction procedure, neighboring MPEs are successively merged, so that an MPE may represent a subsystem containing multiple qubits. 
The tensor-train structure along the temporal direction is preserved throughout the contraction process.

Throughout this work, the term MPE refers both to the elementary tensor trains associated with individual qubits and to the merged tensor trains obtained during the contraction process.
 
Unlike an MPS, which represents a quantum state at a fixed time and organizes correlations across spatial bipartitions, an MPE organizes correlations along the temporal direction. The two representations therefore correspond to different contraction orders of the same underlying two-dimensional tensor network.

\subsection{Encoding a quantum circuit into MPEs }

We next describe how a quantum circuit is encoded in the MPE representation. 

The initial quantum state is represented as an MPS along the spatial direction. 
The local tensor of this initial MPS at each qubit is incorporated into the first time slice $T_0$ of the corresponding MPE.

Single-qubit gates are represented as local tensors and inserted into the MPE of the qubit on which they act. 
If no operation is applied to a qubit at a given time step, an identity tensor is inserted so that the tensor network retains a rectangular structure. 
This construction allows all MPEs to share a common temporal indexing where operations acting on different qubits are applied simultaneously.

Two- and multi-qubit gates are represented as matrix product operators (MPOs). 
The local tensors of an MPO are absorbed into the MPEs of the qubits involved in the gate at the corresponding time step. 
For nearest-neighbor two-qubit gates, the MPO reduces to a two-site representation, and the two resulting local tensors are incorporated into the neighboring MPEs. 
More general multi-qubit gates can be treated in the same manner by decomposing the gate into an MPO and assigning its local tensors to the corresponding MPEs. 
This construction applies regardless of the location or number of qubits involved in each gate.

The resulting representation is a rectangular two-dimensional tensor network whose spatial direction corresponds to the qubit ordering and whose temporal direction corresponds to the circuit depth. 
The rows of this tensor network constitute the elementary MPEs from which the contraction procedure starts.

MPOs are used only to construct this two-dimensional tensor network representation of the circuit. 
The subsequent simulation is performed by compressing and contracting the resulting MPEs. 
In this sense, the MPE representation provides a temporal organization of the quantum circuit, while MPOs provide a convenient local representation of multi-qubit gates within that organization.

\subsection{Temporal compression}

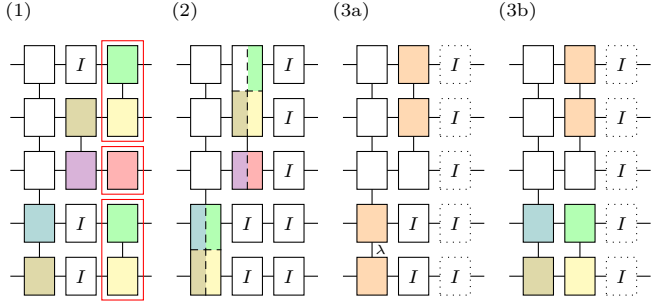
\begin{figure}
\begin{tikzpicture}[T/.style={shape=rectangle, minimum width=0.4cm, minimum height=0.5cm, fill=white, draw},C/.style={shape=rectangle, minimum width=0.4cm, minimum height=1.2cm, fill=white, draw},inner sep=1pt,font=\scriptsize, xscale=0.55, yscale=0.7]
  \node at (-.5,5) {(1)};
  \foreach \y in {0,1,2,3,4} \draw (-.7,\y) -- (2.7,\y);
  \draw (0,0) -- (0,4);
  \draw (1,2) -- (1,3);
  \draw (2,0) -- (2,1);
  \draw (2,3) -- (2,4);

  \node[T,fill=olive!30] at (0,0) {};
  \node[T,fill=teal!30] at (0,1) {};
  \node[T] at (0,2) {};
  \node[T] at (0,3) {};
  \node[T] at (0,4) {};
  \node[T] at (1,0) {$I$};
  \node[T] at (1,1) {$I$};
  \node[T,fill=violet!30] at (1,2) {};
  \node[T,fill=olive!30] at (1,3) {};
  \node[T] at (1,4) {$I$};
  \node[T,fill=yellow!30] at (2,0) {};
  \node[T,fill=green!30] at (2,1) {};
  \node[T,fill=red!30] at (2,2) {};
  \node[T,fill=yellow!30] at (2,3) {};
  \node[T,fill=green!30] at (2,4) {};
  \draw[red] (1.5,-.45) rectangle (2.5,1.45);
  \draw[red] (1.5,1.55) rectangle (2.5,2.45);
  \draw[red] (1.5,2.55) rectangle (2.5,4.45);

  \begin{scope}[shift={(4,0)}]
  \node at (-.5,5) {(2)};
  \foreach \y in {0,1,2,3,4} \draw (-.7,\y) -- (2.7,\y);
  \draw (0,0) -- (0,4);
  \draw (1,2) -- (1,4);

  \node[T,outer sep=0pt,inner sep=0pt,draw=none,fill=yellow!30, anchor=north west, minimum width=0.2cm,minimum height=0.6cm] at (0,.5) {};
  \node[T,outer sep=0pt,inner sep=0pt,draw=none,fill=green!30, anchor=south west, minimum width=0.2cm,minimum height=0.6cm] at (0,.5) {};
  \node[T,outer sep=0pt,inner sep=0pt,draw=none,fill=olive!30, anchor=north east, minimum width=0.2cm,minimum height=0.6cm] at (0,.5) {};
  \node[T,outer sep=0pt,inner sep=0pt,draw=none,fill=teal!30, anchor=south east, minimum width=0.2cm,minimum height=0.6cm] at (0,.5) {};
  \node[C,fill=none] (C21) at (0,.5) {};
  \draw[densely dashed] (C21.south) -- (C21.north);
  \draw[densely dashed] (C21.west) -- (C21.east);

  \node[T] at (0,2) {};
  \node[T] at (0,3) {};
  \node[T] at (0,4) {};
  \node[T] at (1,0) {$I$};
  \node[T] at (1,1) {$I$};
  \node[T,outer sep=0pt,inner sep=0pt,draw=none,fill=red!30, anchor=west, minimum width=0.2cm] at (1,2) {};
  \node[T,outer sep=0pt,inner sep=0pt,draw=none,fill=violet!30, anchor=east, minimum width=0.2cm] at (1,2) {};
  \node[T,fill=none] (T21) at (1,2) {};
  \draw[densely dashed] (T21.south) -- (T21.north);

  \node[T,outer sep=0pt,inner sep=0pt,draw=none,fill=yellow!30, anchor=north west, minimum width=0.2cm,minimum height=0.6cm] at (1,3.5) {};
  \node[T,outer sep=0pt,inner sep=0pt,draw=none,fill=green!30, anchor=south west, minimum width=0.2cm,minimum height=0.6cm] at (1,3.5) {};
  \node[T,outer sep=0pt,inner sep=0pt,draw=none,fill=olive!30, anchor=north east, minimum width=0.2cm,minimum height=0.6cm] at (1,3.5) {};
  \node[T,outer sep=0pt,inner sep=0pt,draw=none, anchor=south east, minimum width=0.2cm,minimum height=0.6cm] at (1,3.5) {};
  \node[C,fill=none] (C22) at (1,3.5) {};
  \draw[densely dashed] (C22.south) -- (C22.north);
  \draw[densely dashed] (C22.west) -- (C22.east);
  \node[T] at (2,0) {$I$};
  \node[T] at (2,1) {$I$};
  \node[T] at (2,2) {$I$};
  \node[T] at (2,3) {$I$};
  \node[T] at (2,4) {$I$};

  \begin{scope}[shift={(4,0)}]
  \node at (-.5,5) {(3a)};

  \foreach \y in {0,1,2,3,4} \draw (-.7,\y) -- (2.7,\y);
  \draw (0,0) -- (0,4);
  \draw (1,2) -- (1,4);
  \node[anchor=west] at (0,.5) {$\scriptstyle\lambda$};

  \node[T,fill=orange!30] at (0,0) {};
  \node[T,fill=orange!30] at (0,1) {};
  \node[T] at (0,2) {};
  \node[T] at (0,3) {};
  \node[T] at (0,4) {};
  \node[T] at (1,0) {$I$};
  \node[T] at (1,1) {$I$};
  \node[T] at (1,2) {};
  \node[T,fill=orange!30] at (1,3) {};
  \node[T,fill=orange!30] at (1,4) {};
  \node[T,dotted] at (2,0) {$I$};
  \node[T,dotted] at (2,1) {$I$};
  \node[T,dotted] at (2,2) {$I$};
  \node[T,dotted] at (2,3) {$I$};
  \node[T,dotted] at (2,4) {$I$};

  \begin{scope}[shift={(4,0)}]
  \node at (-.5,5) {(3b)};

  \foreach \y in {0,1,2,3,4} \draw (-.7,\y) -- (2.7,\y);
  \draw (0,0) -- (0,4);
  \draw (1,0) -- (1,1);
  \draw (1,2) -- (1,4);

  \node[T,fill=olive!30] at (0,0) {};
  \node[T,fill=teal!30] at (0,1) {};
  \node[T] at (0,2) {};
  \node[T] at (0,3) {};
  \node[T] at (0,4) {};
  \node[T,fill=yellow!30] at (1,0) {};
  \node[T,fill=green!30] at (1,1) {};
  \node[T] at (1,2) {};
  \node[T,fill=orange!30] at (1,3) {};
  \node[T,fill=orange!30] at (1,4) {};
  \node[T,dotted] at (2,0) {$I$};
  \node[T,dotted] at (2,1) {$I$};
  \node[T,dotted] at (2,2) {$I$};
  \node[T,dotted] at (2,3) {$I$};
  \node[T,dotted] at (2,4) {$I$};

  \end{scope}
  \end{scope}
  \end{scope}
  \end{tikzpicture}
  \caption[]{Procedure of the temporal compression. 
(1) Each red box indicates a local tensor block to be processed.
(2) The block is contracted with the nearest nontrivial tensors on its left (earlier time step). 
(3a) The result is factorized by an SVD along the spatial direction; if the spatial bond dimension $\lambda$ is at most four, the merge is accepted and the all-identity columns (dotted) vacated by the merge are removed.
(3b) If it exceeds four, the merge is rejected and the block is packed leftward past the identity tensors.}
 
\label{fig:temporal_compression}
\end{figure}

Before contracting the MPEs along the spatial direction, we reduce the effective depth of the tensor network [Fig.~\ref{fig:overview}(2)] to a shorter rectangular tensor network [Fig.~\ref{fig:overview}(3)] by compressing it along the temporal direction. 
The purpose of this procedure is to reduce the effective circuit depth while preserving the tensor-network representation of the circuit.

Since the circuit has already been converted into local tensors, each time slice is first partitioned into local tensor blocks. 
A local tensor block is defined as a connected set of neighboring local tensors separated from other blocks by spatial bonds of dimension one [Fig.~\ref{fig:temporal_compression}(1)].
For the circuits considered in this work, the gate operators are represented as MPOs with bond dimensions not exceeding four. 
Consequently, the spatial bond dimensions appearing within a local tensor block are also bounded by four.

The compression proceeds by successively packing nontrivial local tensor blocks toward earlier time steps. 
When a local tensor block is processed, it is moved toward earlier time steps across regions consisting only of identity tensors. 
Whenever the block encounters nontrivial local tensors, candidate mergers are examined by contracting the participating tensors [Fig.~\ref{fig:temporal_compression}(2)] and factorizing the result exactly along the spatial direction.
If the resulting spatial bond dimension does not exceed four, the merged representation is adopted and the absorbed tensors are replaced by identity tensors [Fig.~\ref{fig:temporal_compression}(3a)].
If instead the spatial bond dimension exceeds four, the merge is rejected.
In that case, the block is placed at the time step immediately to the right of the nontrivial local tensors [Fig.~\ref{fig:temporal_compression}(3b)].
Repeating this procedure progressively packs nontrivial tensor blocks into earlier time steps and reduces the temporal extent occupied by nontrivial tensors.

Since no singular values are discarded during this procedure, the temporal compression introduces no additional approximation. Its role is solely to reorganize the tensor network into a shallower form before the subsequent spatial contraction.

After all local tensor blocks have been processed, columns consisting entirely of identity tensors are removed. The resulting tensor network therefore retains the same circuit representation while having a reduced effective depth. This compressed rectangular tensor network is subsequently used as the input to the zip-up contraction procedure.

This preprocessing step reduces the size of the tensor trains entering the zip-up contraction and thereby improves the efficiency of the subsequent spatial contraction.

\subsection{Zip-up contraction of MPEs}

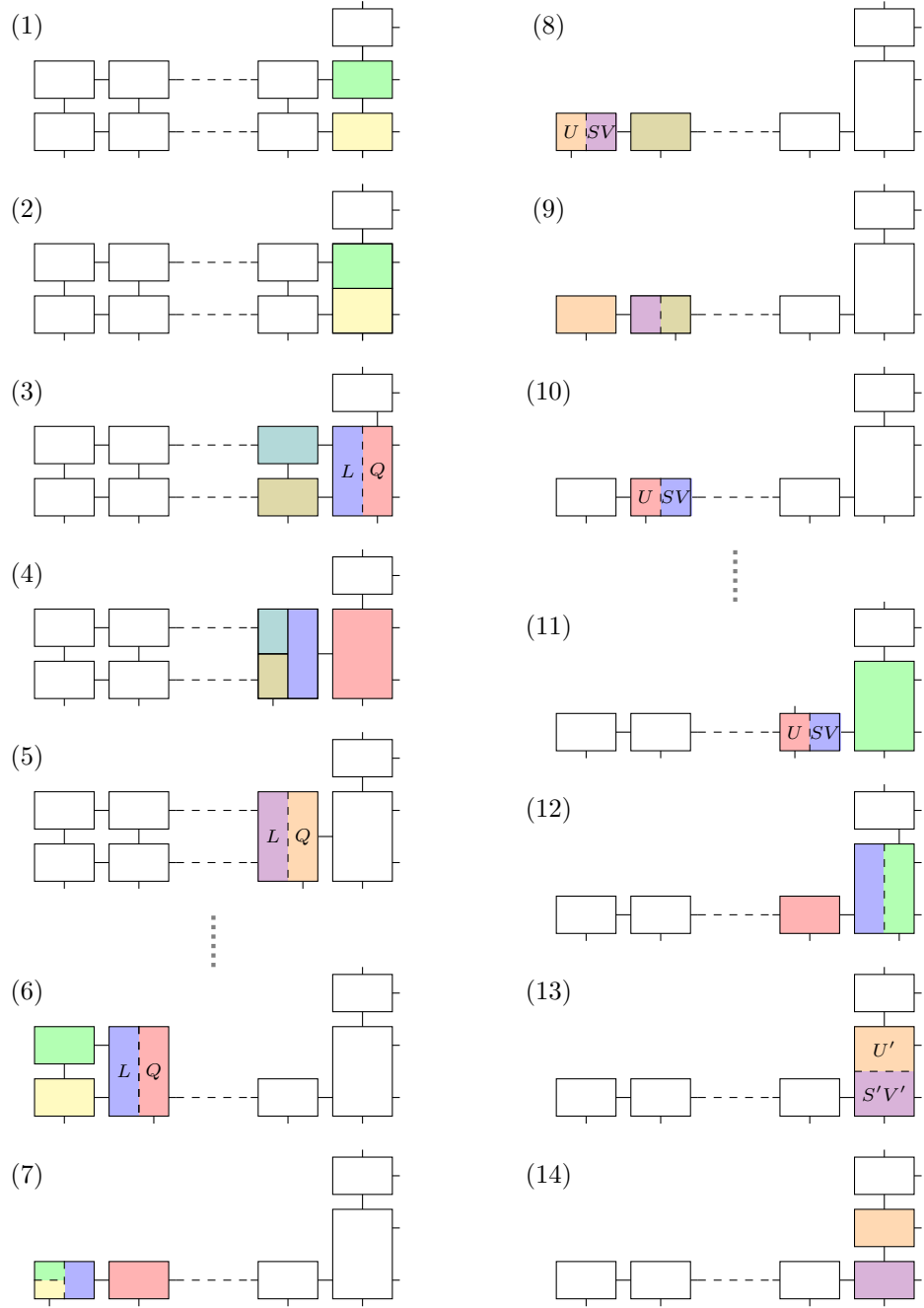
\begin{figure*}[tbph]
\centering
\begin{tikzpicture}[yscale=0.7,T/.style={shape=rectangle, minimum width=0.8cm, minimum height=0.5cm, fill=white, draw},C/.style={shape=rectangle, minimum width=0.8cm, minimum height=1.2cm, fill=white, draw},U/.style={shape=triangle, minimum heght=0}, inner sep=1pt,font=\scriptsize ]
    \node[font=\normalsize] at (.5,3) {(1)};
\draw (1,1) -- (2.5,1);
\draw (1,2) -- (2.5,2);
\draw (3.5,1) -- (5.5,1);
\draw (3.5,2) -- (5.5,2);
\draw[dashed] (2.5,1) -- (3.5,1);
\draw[dashed] (2.5,2) -- (3.5,2);
\draw (5,3) -- (5.5,3);
\draw (1,0.5) -- (1,2);
\draw (2,0.5) -- (2,2);
\draw (4,0.5) -- (4,2);
\draw (5,0.5) -- (5,3.5);
\draw (5,0.5) -- (5,3.5);
\node[T] (T011) at (1,1) {};
\node[T] (T021) at (2,1) {};
\node[T] (T041) at (4,1) {};
\node[T,fill=yellow!30] (T051) at (5,1) {};
\node[T] (T012) at (1,2) {};
\node[T] (T022) at (2,2) {};
\node[T] (T042) at (4,2) {};
\node[T,fill=green!30] (T052) at (5,2) {};
\node[T] (T053) at (5,3) {};

\begin{scope}[shift={(0,-3.5)}]
\node[font=\normalsize] at (.5,3) {(2)};
\draw (1,1) -- (2.5,1);
\draw (1,2) -- (2.5,2);
\draw (3.5,1) -- (5.5,1);
\draw (3.5,2) -- (5.5,2);
\draw[dashed] (2.5,1) -- (3.5,1);
\draw[dashed] (2.5,2) -- (3.5,2);
\draw (5,3) -- (5.5,3);
\draw (1,0.5) -- (1,2);
\draw (2,0.5) -- (2,2);
\draw (4,0.5) -- (4,2);
\draw (5,0.5) -- (5,3.5);
\draw (5,0.5) -- (5,3.5);
\node[T] (T111) at (1,1) {};
\node[T] (T121) at (2,1) {};
\node[T] (T141) at (4,1) {};
\node[C,fill=green!30,minimum height=0.6cm,anchor=south,outer sep=0pt,inner sep=0pt] at (5,1.5) {};
\node[C,fill=yellow!30,minimum height=0.6cm,anchor=north,outer sep=0pt,inner sep=0pt] at (5,1.5) {};
\node[C,fill=none] (C151) at (5,1.5) {};
\node[T] (T112) at (1,2) {};
\node[T] (T122) at (2,2) {};
\node[T] (T142) at (4,2) {};
\node[T] (T153) at (5,3) {};

\begin{scope}[shift={(0,-3.5)}]
\node[font=\normalsize] at (.5,3) {(3)};
\draw (1,1) -- (2.5,1);
\draw (1,2) -- (2.5,2);
\draw (3.5,1) -- (5.5,1);
\draw (3.5,2) -- (5.5,2);
\draw[dashed] (2.5,1) -- (3.5,1);
\draw[dashed] (2.5,2) -- (3.5,2);
\draw (5,3) -- (5.5,3);
\draw (1,0.5) -- (1,2);
\draw (2,0.5) -- (2,2);
\draw (4,0.5) -- (4,2);
\draw (5,3) -- (5,3.5);
\draw (5.2,3) -- (5.2,0.5);
\node[T] (T211) at (1,1) {};
\node[T] (T221) at (2,1) {};
\node[T,fill=olive!30] (T241) at (4,1) {};
\node[T] (T212) at (1,2) {};
\node[T] (T222) at (2,2) {};
\node[T,fill=teal!30] (T242) at (4,2) {};
\node[C,draw=none,fill=red!30,minimum width=0.4cm,anchor=west,outer sep=0pt,inner sep=0pt] at (5,1.5) {$Q$};
\node[C,draw=none,fill=blue!30,minimum width=0.4cm,anchor=east,outer sep=0pt,inner sep=0pt] at (5,1.5) {$L$};
\node[C,fill=none] (C251) at (5,1.5) {};
\node[T] (T253) at (5,3) {};
\draw[dashed] (C251.north) -- (C251.south);

\begin{scope}[shift={(0,-3.5)}]
\node[font=\normalsize] at (.5,3) {(4)};
\draw (1,1) -- (2.5,1);
\draw (1,2) -- (2.5,2);
\draw (3.5,1) -- (4,1);
\draw (5,1) -- (5.5,1);
\draw (4,1.5) -- (5,1.5);
\draw (3.5,2) -- (4,2);
\draw (5,2) -- (5.5,2);
\draw[dashed] (2.5,1) -- (3.5,1);
\draw[dashed] (2.5,2) -- (3.5,2);
\draw (5,3) -- (5.5,3);
\draw (1,0.5) -- (1,2);
\draw (2,0.5) -- (2,2);
\draw (4,1.5) -- (4,2);
\draw (3.8,1.5) -- (3.8,0.5);
\draw (5,0.5) -- (5,3.5);
\node[T] (T311) at (1,1) {};
\node[T] (T321) at (2,1) {};
\node[T] (T312) at (1,2) {};
\node[T] (T322) at (2,2) {};
\node[T,fill=olive!30,minimum width=0.4cm,minimum height=0.6cm, anchor=north east,inner sep=0pt,outer sep=0pt] at (4,1.5) {};
\node[T,fill=teal!30,minimum width=0.4cm,minimum height=0.6cm, anchor=south east,inner sep=0pt,outer sep=0pt] at (4,1.5) {};
\node[C,fill=blue!30,minimum width=0.4cm,anchor=west,inner sep=0pt,outer sep=0pt] at (4,1.5) {};
\node[C,fill=none] (C341) at (4,1.5) {};
\node[C,fill=red!30] (C351) at (5,1.5) {};
\node[T] (T353) at (5,3) {};
\draw (C341.west) -- (C341.center);

\begin{scope}[shift={(0,-3.5)}]
\node[font=\normalsize] at (.5,3) {(5)};
\draw (1,1) -- (2.5,1);
\draw (1,2) -- (2.5,2);
\draw (3.5,1) -- (4,1);
\draw (5,1) -- (5.5,1);F
\draw (4,1.5) -- (5,1.5);
\draw (3.5,2) -- (4,2);
\draw (5,2) -- (5.5,2);
\draw[dashed] (2.5,1) -- (3.5,1);
\draw[dashed] (2.5,2) -- (3.5,2);
\draw (5,3) -- (5.5,3);
\draw (1,0.5) -- (1,2);
\draw (2,0.5) -- (2,2);
\draw (4,1.5) -- (4,2);
\draw (5,0.5) -- (5,3.5);
\draw (4.2,1.5) -- (4.2,0.5);
\node[T] (T411) at (1,1) {};
\node[T] (T421) at (2,1) {};
\node[T] (T412) at (1,2) {};
\node[T] (T422) at (2,2) {};
\node[C,draw=none,fill=orange!30,minimum width=0.4cm,anchor=west,outer sep=0pt,inner sep=0pt] at (4,1.5) {$Q$};
\node[C,draw=none,fill=violet!30,minimum width=0.4cm,anchor=east,outer sep=0pt,inner sep=0pt] at (4,1.5) {$L$};
\node[C,fill=none] (C441) at (4,1.5) {};
\draw[dashed] (C441.north) -- (C441.south);
\node[C] (C451) at (5,1.5) {};
\node[T] (T453) at (5,3) {};

\begin{scope}[shift={(0,-4.5)}]
\node[font=\normalsize] at (.5,3) {(6)};
\draw[gray,ultra thick,dotted] (3,3.5)--(3,4.5);
\draw (1,1) -- (2.5,1);
\draw (1,2) -- (2,2);
\draw (3.5,1) -- (5.5,1);
\draw (5,2) -- (5.5,2);
\draw[dashed] (2.5,1) -- (3.5,1);
\draw (5,3) -- (5.5,3);
\draw (1,0.5) -- (1,2);
\draw (4,0.5) -- (4,1);
\draw (5,0.5) -- (5,3.5);
\draw (2.2,1.5) -- (2.2,0.5);
\node[T,fill=yellow!30] (T511) at (1,1) {};
\node[T,fill=green!30] (T512) at (1,2) {};
\node[C,draw=none,fill=red!30,minimum width=0.4cm,anchor=west,outer sep=0pt,inner sep=0pt] at (2,1.5) {$Q$};
\node[C,draw=none,fill=blue!30,minimum width=0.4cm,anchor=east,outer sep=0pt,inner sep=0pt] at (2,1.5) {$L$};
\node[C,fill=none] (C521) at (2,1.5) {};
\draw[dashed] (C521.north) -- (C521.south);
\node[T] (T541) at (4,1) {};
\node[C] (C551) at (5,1.5) {};
\node[T] (T553) at (5,3) {};
\draw[dashed] (C521.north) -- (C521.south);

\begin{scope}[shift={(0,-3.5)}]
\node[font=\normalsize] at (.5,3) {(7)};
\draw (1,1) -- (2.5,1);
\draw (3.5,1) -- (5.5,1);
\draw (5,2) -- (5.5,2);
\draw[dashed] (2.5,1) -- (3.5,1);
\draw (5,3) -- (5.5,3);
\draw (2,0.5) -- (2,1);
\draw (4,0.5) -- (4,1);
\draw (5,0.5) -- (5,3.5);
\draw (0.8,1) -- (0.8,0.5);
\node[T,draw=none,fill=blue!30,minimum width=0.4cm,anchor=west,outer sep=0pt,inner sep=0pt] at (1,1) {};
\node[T,draw=none,fill=green!30,minimum width=0.4cm, minimum height=0.25cm,outer sep=0pt,inner sep=0pt,anchor =south east] at (1,1) {};
\node[T,draw=none,fill=yellow!30,minimum width=0.4cm, minimum height=0.25cm,outer sep=0pt,inner sep=0pt,anchor =north east] at (1,1) {};
\node[T,fill=none] (C611) at (1,1) {};
\node[T,fill=red!30] (C621) at (2,1) {};
\node[T] (C641) at (4,1) {};
\node[C] (C651) at (5,1.5) {};
\node[T] (T653) at (5,3) {};
\draw[dashed] (C611.north) -- (C611.south);
\draw[dashed] (C611.west) -- (C611.center);
\end{scope}
\end{scope}
\end{scope}
\end{scope}
\end{scope}
\end{scope}

\begin{scope}[shift={(7,0)}]
\node[font=\normalsize] at (.5,3) {(8)};
\draw (1,1) -- (2.5,1);
\draw (3.5,1) -- (5.5,1);
\draw (5,2) -- (5.5,2);
\draw[dashed] (2.5,1) -- (3.5,1);
\draw (5,3) -- (5.5,3);
\draw (2,0.5) -- (2,1);
\draw (4,0.5) -- (4,1);
\draw (5,0.5) -- (5,3.5);
\draw (0.8,1) -- (0.8,0.5);
\node[T,fill=olive!30] (T121) at (2,1) {};
\node[T] (T141) at (4,1) {};
\node[C] (T151) at (5,1.5) {};
\node[T] (T153) at (5,3) {};
\node[T,draw=none,fill=orange!30,minimum width=0.4cm,,outer sep=0pt,inner sep=0pt,anchor=east] at (1,1) {$U$};
\node[T,draw=none,fill=violet!30,minimum width=0.4cm,,outer sep=0pt,inner sep=0pt,anchor=west] at (1,1) {$SV$};
\node[T,fill=none] (T111) at (1,1) {};
\draw[dashed] (T111.north) -- (T111.south);

\begin{scope}[shift={(0,-3.5)}]
\node[font=\normalsize] at (.5,3) {(9)};
\draw (1,1) -- (2.5,1);
\draw (3.5,1) -- (5.5,1);
\draw[dashed] (2.5,1) -- (3.5,1);
\draw (5,3) -- (5.5,3);
\draw (1,0.5) -- (1,1);
\draw (4,0.5) -- (4,1);
\draw (5,0.5) -- (5,3.5);
\draw (2.2,1) -- (2.2,0.5);
\node[T,fill=orange!30] (T211) at (1,1) {};
\node[T] (T221) at (2,1) {};
\node[T] (T241) at (4,1) {};
\node[C] (C251) at (5,1.5) {};
\node[T] (T253) at (5,3) {};
\node[T,draw=none,fill=violet!30,minimum width=0.4cm,,outer sep=0pt,inner sep=0pt,anchor=east] at (2,1) {};
\node[T,draw=none,fill=olive!30,minimum width=0.4cm,,outer sep=0pt,inner sep=0pt,anchor=west] at (2,1) {};
\node[T,fill=none] (T221) at (2,1) {};
\draw[dashed] (T221.north) -- (T221.south);

\begin{scope}[shift={(0,-3.5)}]
\node[font=\normalsize] at (.5,3) {(10)};
\draw (1,1) -- (2.5,1);
\draw (3.5,1) -- (5.5,1);
\draw (5,2) -- (5.5,2);
\draw[dashed] (2.5,1) -- (3.5,1);
\draw (5,3) -- (5.5,3);
\draw (1,0.5) -- (1,1);
\draw (4,0.5) -- (4,1);
\draw (5,0.5) -- (5,3.5);
\draw (1.8,1) -- (1.8,0.5);
\node[T] (T311) at (1,1) {};
\node[T] (T341) at (4,1) {};
\node[C] (C351) at (5,1.5) {};
\node[T] (T353) at (5,3) {};
\node[T,draw=none,fill=red!30,minimum width=0.4cm,,outer sep=0pt,inner sep=0pt,anchor=east] at (2,1) {$U$};
\node[T,draw=none,fill=blue!30,minimum width=0.4cm,,outer sep=0pt,inner sep=0pt,anchor=west] at (2,1) {$SV$};
\node[T,fill=none] (T321) at (2,1) {};
\draw[dashed] (T321.north) -- (T321.south);

\begin{scope}[shift={(0,-4.5)}]
\node[font=\normalsize] at (.5,3) {(11)};
\draw[gray,ultra thick,dotted] (3,3.5)--(3,4.5);
\draw (1,1) -- (2.5,1);
\draw (3.5,1) -- (5.5,1);
\draw (5,2) -- (5.5,2);
\draw[dashed] (2.5,1) -- (3.5,1);
\draw (5,3) -- (5.5,3);
\draw (1,0.5) -- (1,1);
\draw (2,0.5) -- (2,1);
\draw (5,0.5) -- (5,3.5);
\draw (3.8,1.5) -- (3.8,0.5);
\node[T] (C411) at (1,1) {};
\node[T] (C421) at (2,1) {};
\node[C,fill=green!30] (C451) at (5,1.5) {};
\node[T] (T453) at (5,3) {};
\node[T,draw=none,fill=red!30,minimum width=0.4cm,,outer sep=0pt,inner sep=0pt,anchor=east] at (4,1) {$U$};
\node[T,draw=none,fill=blue!30,minimum width=0.4cm,,outer sep=0pt,inner sep=0pt,anchor=west] at (4,1) {$SV$};
\node[T,fill=none] (T441) at (4,1) {};
\draw[dashed] (T441.north) -- (T441.south);

\begin{scope}[shift={(0,-3.5)}]
\node[font=\normalsize] at (.5,3) {(12)};
\draw (1,1) -- (2.5,1);
\draw (3.5,1) -- (5.5,1);
\draw (5,2) -- (5.5,2);
\draw[dashed] (2.5,1) -- (3.5,1);
\draw (5,3) -- (5.5,3);
\draw (1,0.5) -- (1,1);
\draw (2,0.5) -- (2,1);
\draw (4,0.5) -- (4,1);
\draw (5,3) -- (5,3.5);
\draw (5.2,3) -- (5.2,0.5);
\node[T] (C511) at (1,1) {};
\node[T] (C521) at (2,1) {};
\node[T,fill=red!30] (C541) at (4,1) {};
\node[T] (T553) at (5,3) {};
\node[C,draw=none,fill=blue!30,minimum width=0.4cm,,outer sep=0pt,inner sep=0pt,anchor=east] at (5,1.5) {};
\node[C,draw=none,fill=green!30,minimum width=0.4cm,,outer sep=0pt,inner sep=0pt,anchor=west] at (5,1.5) {};
\node[C,fill=none] (C551) at (5,1.5) {};
\draw[dashed] (C551.north) -- (C551.south);

\begin{scope}[shift={(0,-3.5)}]
\node[font=\normalsize] at (.5,3) {(13)};
\draw (1,1) -- (2.5,1);
\draw (3.5,1) -- (5.5,1);
\draw (5,2) -- (5.5,2);
\draw[dashed] (2.5,1) -- (3.5,1);
\draw (5,3) -- (5.5,3);
\draw (1,0.5) -- (1,1);
\draw (2,0.5) -- (2,1);
\draw (4,0.5) -- (4,1);
\draw (5,0.5) -- (5,3.5);
\node[T] (C611) at (1,1) {};
\node[T] (C621) at (2,1) {};
\node[T] (C641) at (4,1) {};
\node[T,fill=none,draw=none] (T651) at (5,1) {SV};
\node[T,fill=none,draw=none] (T652) at (5,2) {U};
\node[T] (T653) at (5,3) {};
\node[C,draw=none,fill=orange!30,minimum height=0.6cm,,outer sep=0pt,inner sep=0pt,anchor=south] at (5,1.5) {$U'$};
\node[C,draw=none,fill=violet!30,minimum height=0.6cm,,outer sep=0pt,inner sep=0pt,anchor=north] at (5,1.5) {$S'V'$};
\node[C,fill=none] (C651) at (5,1.5) {};
\draw[dashed] (C651.west) -- (C651.east);

\begin{scope}[shift={(0,-3.5)}]
\node[font=\normalsize] at (.5,3) {(14)};
\draw (1,1) -- (2.5,1);
\draw (3.5,1) -- (5.5,1);
\draw (5,2) -- (5.5,2);
\draw[dashed] (2.5,1) -- (3.5,1);
\draw (5,3) -- (5.5,3);
\draw (1,0.5) -- (1,1);
\draw (2,0.5) -- (2,1);
\draw (4,0.5) -- (4,1);
\draw (5,0.5) -- (5,3.5);
\node[T] (C711) at (1,1) {};
\node[T] (C721) at (2,1) {};
\node[T] (C741) at (4,1) {};
\node[T,fill=violet!30] (T751) at (5,1) {};
\node[T,fill=orange!30] (T752) at (5,2) {};
\node[T] (T753) at (5,3) {};
\end{scope}
\end{scope}
\end{scope}
\end{scope}
\end{scope}
\end{scope}
\end{scope}
\end{tikzpicture}
\caption{Procedure for zipping up two MPEs. The process consists of three parts. In (1)–(7), tensors in adjacent MPEs are contracted, and the resulting lower-triangular matrix $L$ is propagated toward the left. In (8)–(12), singular value decomposition is performed, and the singular values $S$ together with the right unitary matrix $V$ are propagated toward the right. In (13)–(14), an MPS representation of the final state is constructed.}
\label{fig:zipup}
\end{figure*}

The full contraction of the circuit is performed by successively merging neighboring MPEs along the spatial direction. 
A naive contraction of two MPEs would rapidly increase the temporal bond dimensions. To control this growth, we employ a zip-up contraction procedure that combines local contractions with canonicalization and singular-value truncation. 
This procedure is inspired by zip-up and related compression techniques commonly used in MPS and MPO calculations \cite{Paeckel2019, Isakov2021, Stoudenmire2010}. 

The zip-up contraction of two neighboring MPEs consists of three stages, as illustrated in Fig.~\ref{fig:zipup}. 
In Fig.~\ref{fig:zipup}(1), the tensor located in the upper-right corner is assumed to be unitary, as it is generated in Fig.~\ref{fig:zipup}(14) of the preceding zip-up contraction.

First, tensors at corresponding time steps are locally contracted [Fig.~\ref{fig:zipup}(2)], and the resulting tensors are decomposed using an LQ factorization [Fig.~\ref{fig:zipup}(3)]. 
The lower-triangular factor generated by this decomposition is propagated toward earlier time steps [Fig.~\ref{fig:zipup}(4)]. 
This backward sweep brings the tensors into a suitable canonical form from the final time step toward the initial time step [Fig.~\ref{fig:zipup}(5)--(7)]. 

Second, a forward sweep is performed from the initial time step to the final time step. 
At each time step, the local tensor is decomposed by singular value decomposition (SVD). 
The singular values and the right unitary matrix are propagated to the next time step [Fig.~\ref{fig:zipup}(8)--(12)]. 
If the number of singular values exceeds a prescribed maximum bond dimension $\chi$, only the largest $\chi$ singular values are retained. 
This truncation controls the growth of temporal bond dimensions during the contraction. 

Finally, the tensor at the last time step is decomposed by SVD [Fig.~\ref{fig:zipup}(13)], and the remaining unitary tensor is used to construct an MPS representation of the output quantum state [Fig.~\ref{fig:zipup}(14)]. 
In this way, the contraction of neighboring MPEs produces a new MPE representing the combined subsystem, while maintaining a controlled temporal bond dimension.

During successive tensor decompositions and propagations of the zip-up procedure, non-unitary contributions are accumulated and encoded in the singular-value spectrum obtained from the SVD.
Because the remaining tensors are brought into canonical form by the backward and forward sweeps, the singular values obtained from each SVD coincide with the Schmidt coefficients of the corresponding
bipartition.
Truncating small singular values therefore yields the optimal low-rank approximation for the corresponding bipartition in the Frobenius norm, consistent with the standard truncation strategy used in MPS and tensor-network algorithms \cite{Verstraete2008, PerezGarcia2007, Orus2014, Orus2019, Fishman2022, Gray2018}.

The contraction of all MPEs can be performed sequentially or in a partitioned manner. 
In the sequential approach, MPEs are merged one by one along the spatial direction. 
Alternatively, the set of MPEs can be divided into groups, each of which is contracted independently before a final merging step. 
In this work, we use a partition into two groups along the spatial direction and contract the network from both sides before performing the final merge. 
This choice balances computational efficiency and implementation complexity. 
The tensor decompositions, canonicalization procedures, and bond-dimension truncations used here follow standard tensor-network techniques employed in modern MPS and MPO computations and their software implementations \cite{Fishman2022, Gray2018}.

\subsection{Bond-dimension bounds and computational cost}

We now discuss the growth of temporal bond dimensions before truncation.    
After merging $j$ qubits, the Hilbert-space dimension of the merged subsystem is $2^j$. 
This gives an upper bound on the rank of any bipartition within the merged subsystem. 

For an MPE constructed from a product initial state such as $|0\rangle^{\otimes n}$, the bond dimension of the internal leg between $T_{k-1}$ and $T_k$ is bounded by 
\[ 
\min \{4^k,\, 4^{d-k-1} 2^j,\, 2^j\} = \min \{4^k,\, 2^j\}. 
\] 
Here, $4^k$ and $4^{d-k-1}2^j$ represent upper bounds arising from the two sides of the bipartition, while $2^j$ is the Hilbert-space dimension of the merged subsystem.

When the initial state is highly entangled, the initial MPS may already contain large spatial bond dimensions $2^j$. 
Since these MPS bonds are incorporated into the MPE construction, this large bond dimension is inherited by the temporal direction. 
In this case, the bond dimension between $T_{k-1}$ and $T_k$ is bounded by
 \[ 
\min \{4^k 2^j,\, 4^{d-k-1} 2^j,\, 2^j\} = 2^j. 
\]

 These bounds describe the maximum bond dimensions that may appear before truncation. 
 In practical simulations, the bond dimension is restricted to a prescribed maximum value $\chi$ by SVD truncation during the zip-up procedure. 
 Consequently, each local tensor decomposition or factorization acts on tensor dimensions of order $\chi$ and therefore scales as $O(\chi^3)$. 

A zip-up sweep over an MPE of depth $d$ requires $O(d\chi^3)$ operations. 
The contraction of two neighboring MPEs therefore scales as $O(d\chi^3)$, while the sequential contraction of $n$ MPEs scales as $O(nd\chi^3)$. 
The corresponding memory requirement scales as $O(d\chi^2)$. 



 The parameter $\chi$ thus controls the trade-off between computational cost and approximation accuracy. Larger values of $\chi$ allow more temporal correlations to be retained but increase both the computational time and the memory consumption.

\subsection{Effect of post-selection}

We finally consider the effect of post-selection on the MPE contraction. 
Post-selection is a standard conditioning procedure in quantum computation and plays an important role in post-selected and measurement-based protocols \cite{Bayraktar2023, Aaronson2005, Raussendorf2001, Briegel2001, Nielsen2006, Browne2006, Jozsa2006}.

Suppose that, after $j$ qubits have been merged, post-selection is imposed on $p$ of them. 
Post-selection fixes the measurement outcomes of selected qubits and thereby reduces the effective degrees of freedom contributing to the remaining tensor network. 

For a product initial state, the corresponding bond-dimension bound becomes 
\[ 
\min \{4^k,\, 4^{d-k-1} 2^{j-p},\, 2^j\}. 
\] 

For a highly entangled initial state, the bound becomes 
\[ 
\min \{4^k 2^j,\, 4^{d-k-1} 2^{j-p},\, 2^j\} = \min \{4^{d-k-1}2^{j-p},\, 2^j\}. 
\] 

Compared with the bounds without post-selection, the factor $2^j$ in the second term is replaced by $2^{j-p}$. 
Therefore, whenever this term determines the minimum, post-selection reduces the bond-dimension bound by a factor of $2^{-p}$.

This reduction is a direct consequence of the temporal organization of the MPE representation. 
By fixing the final state of selected qubits, post-selection removes part of the temporal degrees of freedom that would otherwise contribute to the MPE contraction. 
As a result, the effective temporal bond dimensions can be reduced, leading to a lower truncation error for a fixed maximum bond dimension $\chi$. 

In contrast, post-selection does not impose an equivalent constraint on the spatial bond dimensions that appear in an MPS representation. 
The effect of post-selection is therefore fundamentally different in MPE- and MPS-based simulations.
This mechanism plays an important role in the numerical results discussed in the following section.

\section{Results}

We apply the MPE-based method to two types of 14 qubit quantum circuits and compare the results with those obtained using the MPS-based approach. 
The target circuits are as follows: (i) circuits composed of randomly generated two-qubit unitary gates, and (ii) circuits representing the quantum Ising model.

To evaluate the accuracy of the truncated simulations, we compute the fidelity between the approximate final state $|\varphi\rangle$ and the exact state $|\varphi_0\rangle$.
Fidelity is defined as $|\langle\varphi|\varphi_0\rangle|^2$, and the corresponding infidelity $1 - |\langle \varphi | \varphi_0 \rangle|^2$ is used as an error measure throughout this section. 
This definition allows us to quantify the loss of information introduced by bond-dimension truncation in both the MPS-based and MPE-based simulations in a unified manner.

For each type of circuit acting on $n$ qubits, we consider two classes of initial states: (a) the product state $|0\rangle^{\otimes n}$ and (b) highly entangled states. 
In the latter case, the initial state is generated as a random state in the $2^n$-dimensional Hilbert space, where each component is assigned a magnitude drawn independently of a uniform distribution in $[0,1]$ and a phase drawn independently of a uniform distribution in $(0,2\pi)$. 
The resulting state vector is normalized and converted into an MPS, whose spatial bond dimensions are truncated to a fixed upper limit. 
After truncation, the MPS is normalized to unit norm, mimicking situations in which the bond dimension has already saturated during an MPS-based simulation.
We also examine the effect of post-selection applied to a subset of qubits in the final state and analyze how it modifies the approximation accuracy in both methods.

\subsection{Randomly-generated gates}

\begin{figure}[tb]
    \centering
    \begin{tikzpicture}[yscale=-.5,xscale=1,font=\scriptsize]
    \foreach \y in {0,1,2,3,4,5,6}
       \draw (0,\y) -- (7,\y) node[at start,left]  {$|0\rangle$};
    \foreach \x in {0,2,4}
    {
    \foreach [evaluate=\i as \j using int(\i+\x*3)] \y/\i in {0/1, 2/2, 4/3} 
    {\draw[fill=white] (\x+0.7,\y-0.3) rectangle (\x+1.3, \y+1.3);
     \node at (\x+1,\y+.5) {$U_{\j}$};
     }
    \foreach [evaluate=\i as \j using int(\i+\x*3)] \y/\i in {1/4, 3/5, 5/6}
     {\draw[fill=white] (\x+1.7,\y-0.3) rectangle (\x+2.3, \y+1.3);
     \node at (\x+2,\y+.5) {$U_{\j}$};
     }
    } 
\end{tikzpicture}
    \caption{Example of a quantum circuit composed of randomly-generated unitary gates $U_j$. 
    This example illustrates a 7-qubit circuit with depth 6.}
    \label{fig:rqc}
\end{figure}
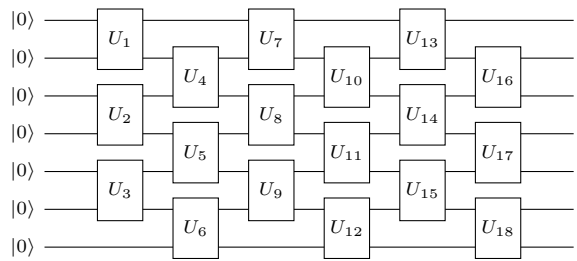

\begin{figure}[tb]
\includegraphics[width=\linewidth]{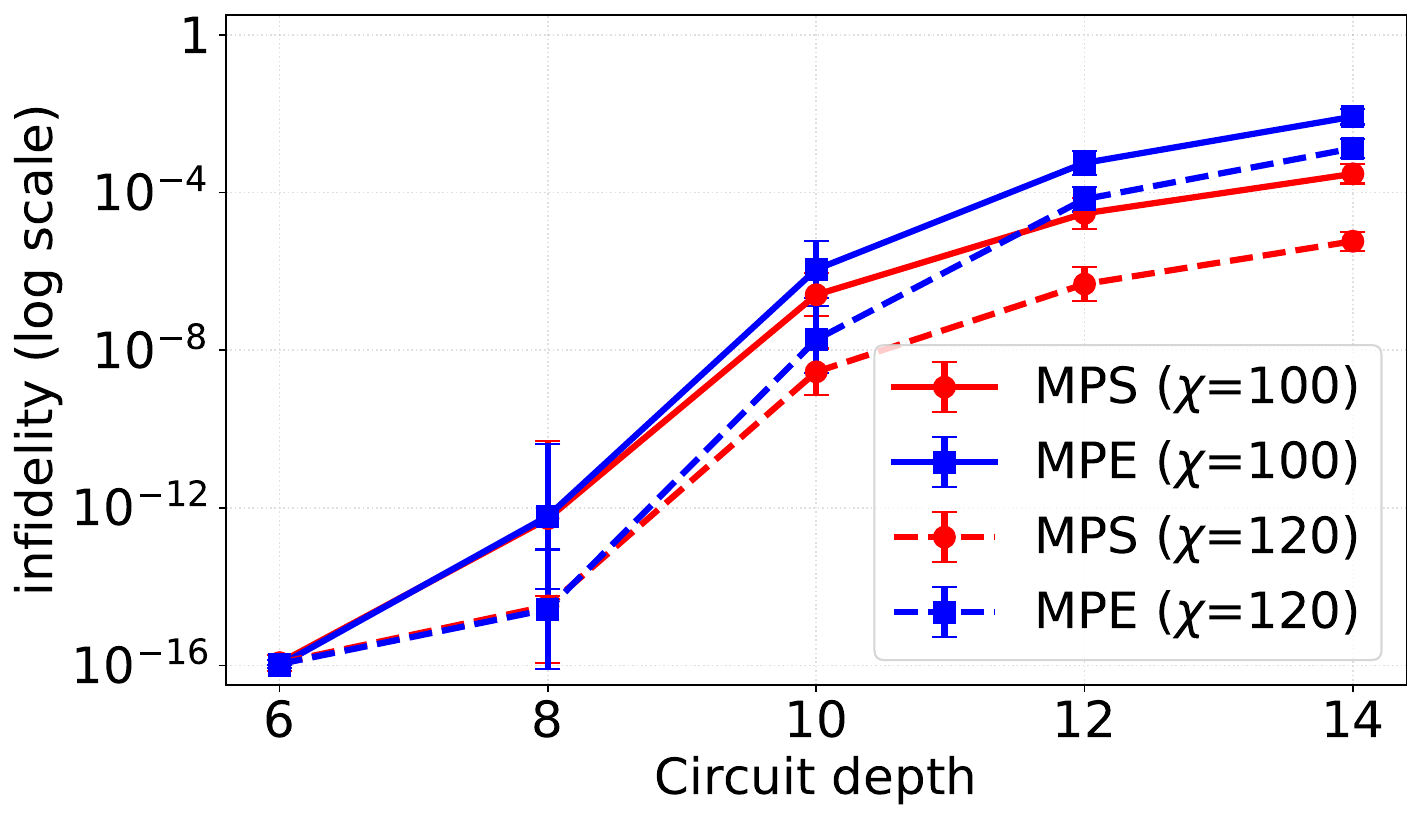}
\caption{Approximation errors of the MPS- and MPE-based methods
for random quantum circuits on 14 qubits with maximum bond
dimensions $\chi=120$ and $100$, compared with the exact results.
The horizontal axis represents the circuit depth. The plotted values
show the mean over $10$ independent trials, and the error bars indicate
the corresponding variance. The initial state is fixed to the product
state $|0\rangle^{\otimes n}$, and no post-selection is applied.
Values below $10^{-16}$ are within machine precision and can be regarded as effectively zero.}
\label{fig:rqc-fidelity}
\end{figure}

\begin{figure}[t]
    \includegraphics[width=\linewidth]{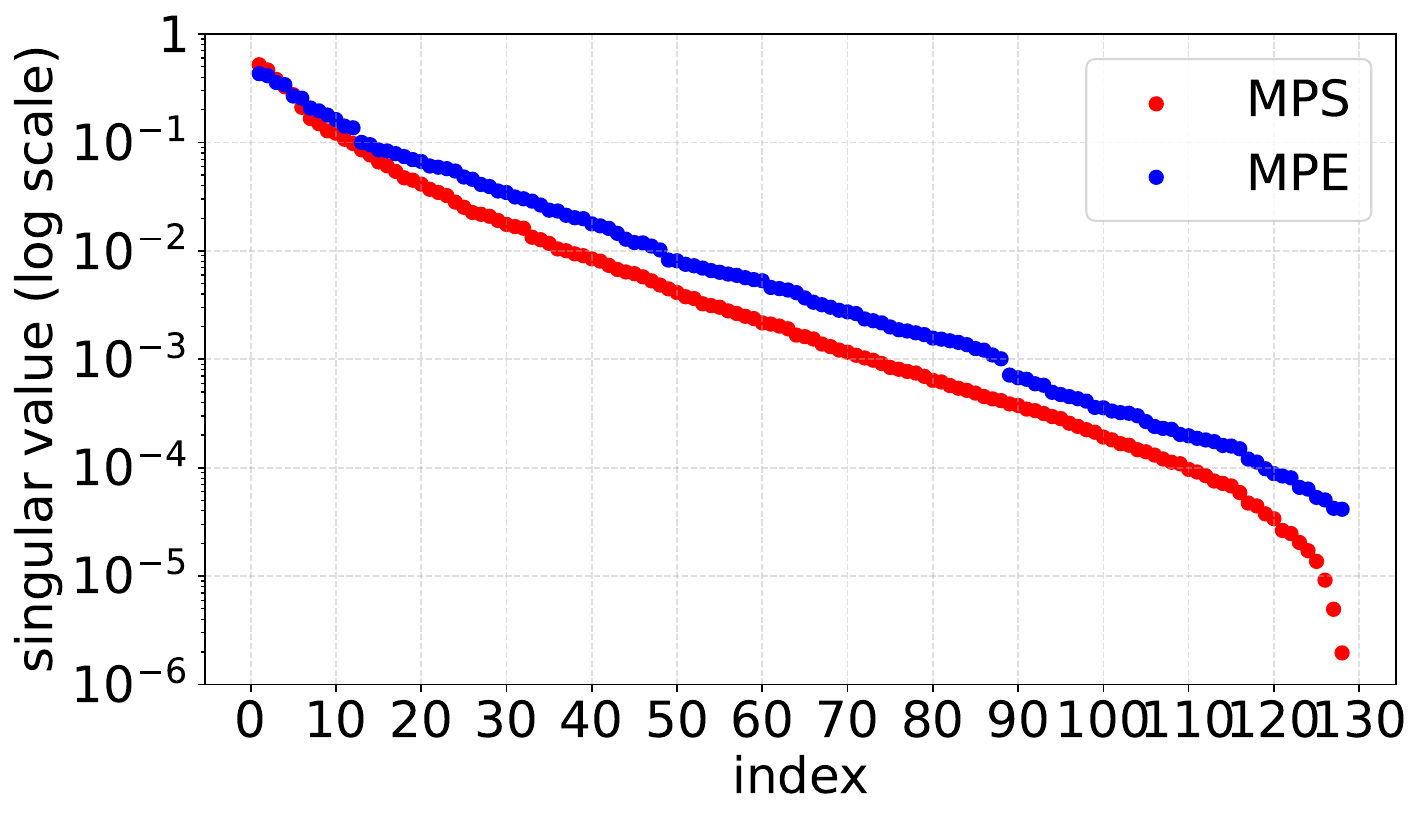}
    \caption{Example of the distribution of singular values while performing the MPS-based and the MPE-based simulations for a randomly generated circuit of 14 qubits with $|0\rangle^{\otimes n}$ as the initial state. 
    The singular values are normalized such that the sum of their squares equals one.}
    \label{fig:rqc-singular-value}
\end{figure}

We first consider quantum circuits composed of randomly generated two-qubit unitary gates, as shown in Fig.~\ref{fig:rqc}, acting on $14$ qubits, with the initial state given by the product state $|0\rangle^{\otimes n}$. 
Random quantum circuits are widely used as benchmarks for classical simulation and quantum-device validation \cite{Boixo2018, Arute2019, Markov2018, Villalonga2019, Huang2020, Pan2022, Haener2017, Smelyanskiy2016}.
Each unitary matrix for the two-qubit gate is generated randomly as follows.
Each element is assigned a complex value, with its magnitude drawn from a uniform distribution on $[0, 1]$ and its phase drawn uniformly from $[0, 2\pi]$. 
The resulting matrix is then converted to a unitary matrix by QR decomposition.

We focus on bond dimensions $\chi = 100$ and $120$, for which both methods still provide meaningful approximations over a nontrivial range of circuit depths.
Figure~\ref{fig:rqc-fidelity} compares the approximation accuracy of MPS- and MPE-based methods under bond-dimension truncation. 
Both methods reproduce the exact results when the bond dimensions remain below the truncation threshold $\chi$, so that singular values are not discarded.
However, once truncation becomes active, the approximation error of the MPE-based method increases more rapidly than that of the MPS-based method. 
This behavior can be understood from the singular-value distributions shown in Fig.~\ref{fig:rqc-singular-value}.
Here, the singular values are taken from the bond with the largest dimension in the final MPS or MPE obtained after the contraction. 
The singular values are normalized so that the sum of their squares is equal to unity.
The singular-value distributions reveal a clear difference between the two methods. 
In the MPS-based simulation, the singular values decay rapidly and become negligible at larger indices, implying that the discarded contributions are already insignificant. 
By contrast, the singular values in the MPE-based simulation remain comparatively large even at higher indices. 
Consequently, truncation removes more important contributions, resulting in a larger loss of information and a faster growth of the approximation error.

\begin{figure}[t]
\includegraphics[width=\linewidth]{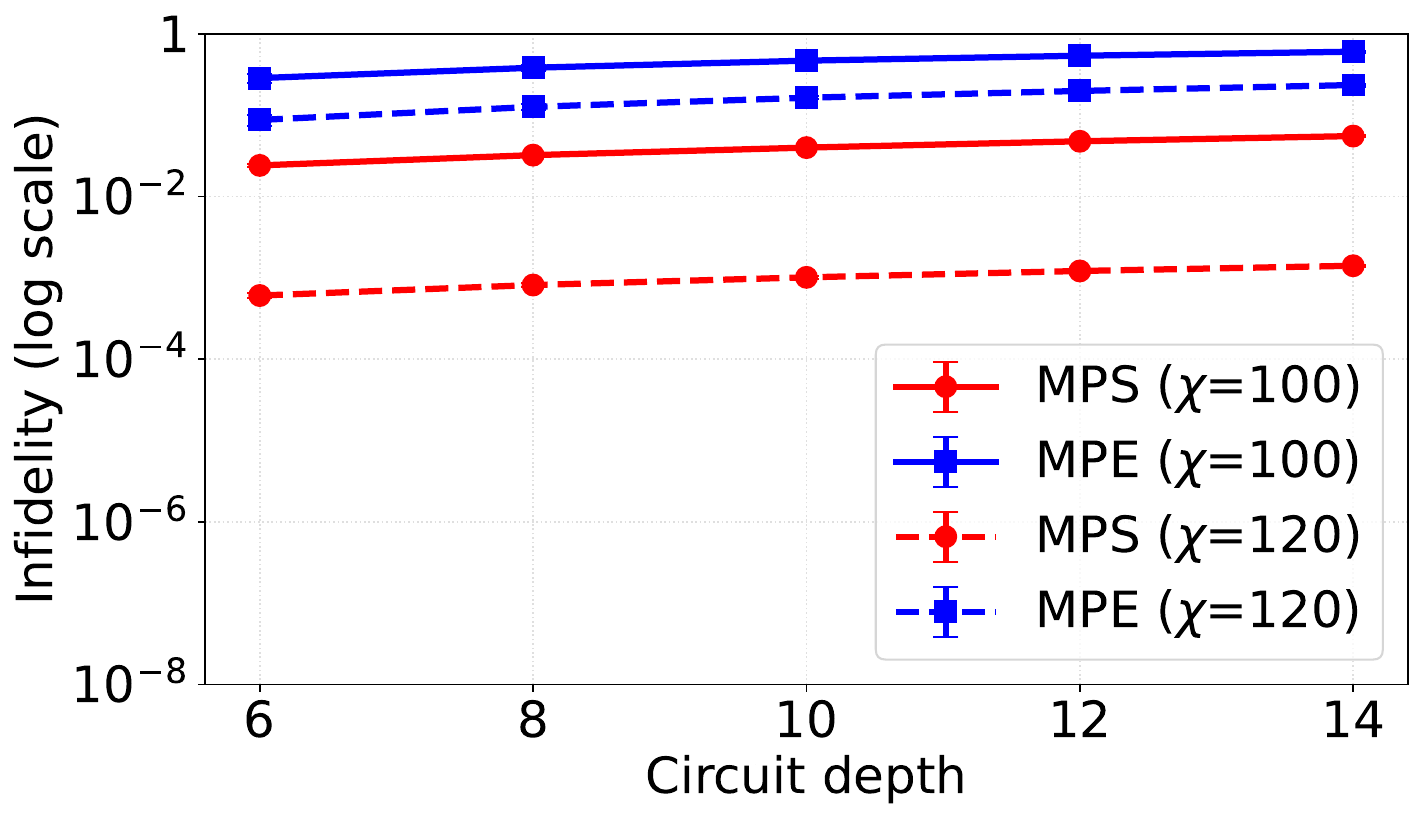}
    \caption{Approximation errors of the MPS- and MPE-based methods for random quantum circuits on $14$ qubits with a highly entangled initial state. The other conditions are the same as in Fig.~\ref{fig:rqc-fidelity}.}
    \label{fig:rqc-random-state-fidelity}
\end{figure}

We also investigate the case where the initial state is highly entangled using randomly generated quantum states. 
Although highly entangled initial states substantially affect the behavior of MPS-based simulations, the results in Fig.~\ref{fig:rqc-random-state-fidelity} indicate that the MPE-based method does not exhibit a clear advantage in accuracy over the MPS-based approach in this setting. 

\begin{figure}[t]
\includegraphics[width=\linewidth]{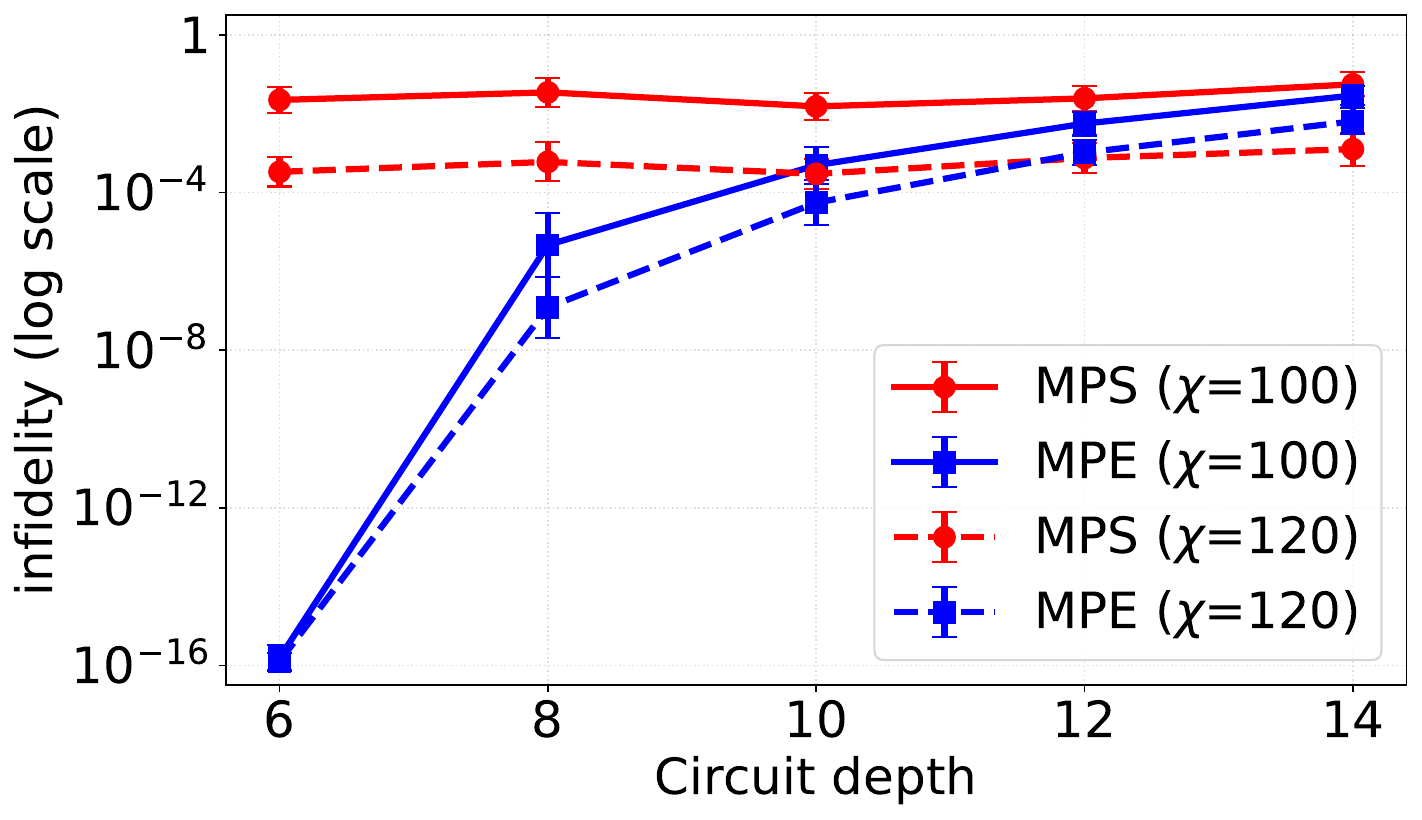}
    \caption{Approximation errors of the MPS- and MPE-based methods for random quantum circuits on $14$ qubits with a highly entangled initial state, where post-selection is applied to all qubits except the boundary qubits. The other conditions are the same as in Fig.~\ref{fig:rqc-random-state-fidelity}.}
    \label{fig:rqc-random-state-post-selection-fidelity}
\end{figure}

This situation changes markedly when post-selection is introduced.
As shown in Fig.~\ref{fig:rqc-random-state-post-selection-fidelity}, post-selection substantially reduces the infidelity of the MPE-based method for highly entangled initial states, whereas the MPS-based method exhibits only a limited improvement.
In the MPE-based framework, post-selection reduces the effective bond dimension by eliminating degrees of freedom associated with the measured qubits, which directly improves the accuracy of truncation. 
In contrast, in the MPS-based approach, post-selection does not reduce the bond dimension in the same manner, and therefore does not lead to a comparable improvement in approximation accuracy. 
As a result, even in regimes where the bond dimension would otherwise grow rapidly, the MPE-based method maintains higher accuracy when post-selection is applied.
These observations indicate that the principal advantage of the MPE-based method in this setting originates from the reduction of temporal degrees of freedom induced by post-selection.

\subsection{Quantum Ising model}
\begin{figure}[t]
    \centering
    \begin{tikzpicture}[yscale=-.5,xscale=1.3,font=\scriptsize]
    \foreach \y in {0,1,2,3,4,5,6}
       \draw (0.5,\y) -- (7,\y); 
    \foreach \y/\i in {0/1,2/2,4/3}
    {
      \draw (1,\y) -- +(0,1) node [fill=white,draw,circle,inner sep=0pt] {+}; 
      \node[draw,fill=black,circle,inner sep=1pt] at (1,\y) {};
      \node [fill=white,draw] at (1.75,\y+1) {$R_z(\theta_{\i})$};
      \draw (2.5,\y) -- +(0,1) node [fill=white,draw,circle,inner sep=0pt] {+}; 
      \node[draw,fill=black,circle,inner sep=1pt] at (2.5,\y) {};
    }
    \foreach \y/\i in {1/4,3/5,5/6}
    {
      \draw (3,\y) -- +(0,1) node [fill=white,draw,circle,inner sep=0pt] {+}; 
      \node[draw,fill=black,circle,inner sep=1pt] at (3,\y) {};
      \node [fill=white,draw] at (3.75,\y+1) {$R_z(\theta_{\i})$};
      \draw (4.5,\y) -- +(0,1) node [fill=white,draw,circle,inner sep=0pt] {+}; 
      \node[draw,fill=black,circle,inner sep=1pt] at (4.5,\y) {};
    }
    \foreach \y in {0,1,2,3,4,5,6}
    {
      \node [fill=white,draw] at (5.25,\y) {$R_x(\varphi)$};
      \node [fill=white,draw] at (6.25,\y) {$R_z(\varphi_{\y})$};
    }
\end{tikzpicture}
    \caption{A single Trotter step of the quantum circuit used to simulate the time evolution of the Ising model with $n=7$.}
    \label{fig:ising}
\end{figure}
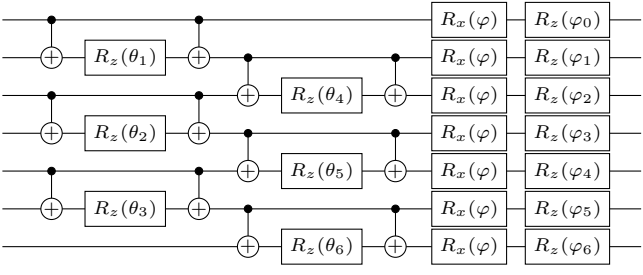

\begin{figure}[t]
\includegraphics[width=0.48\textwidth]{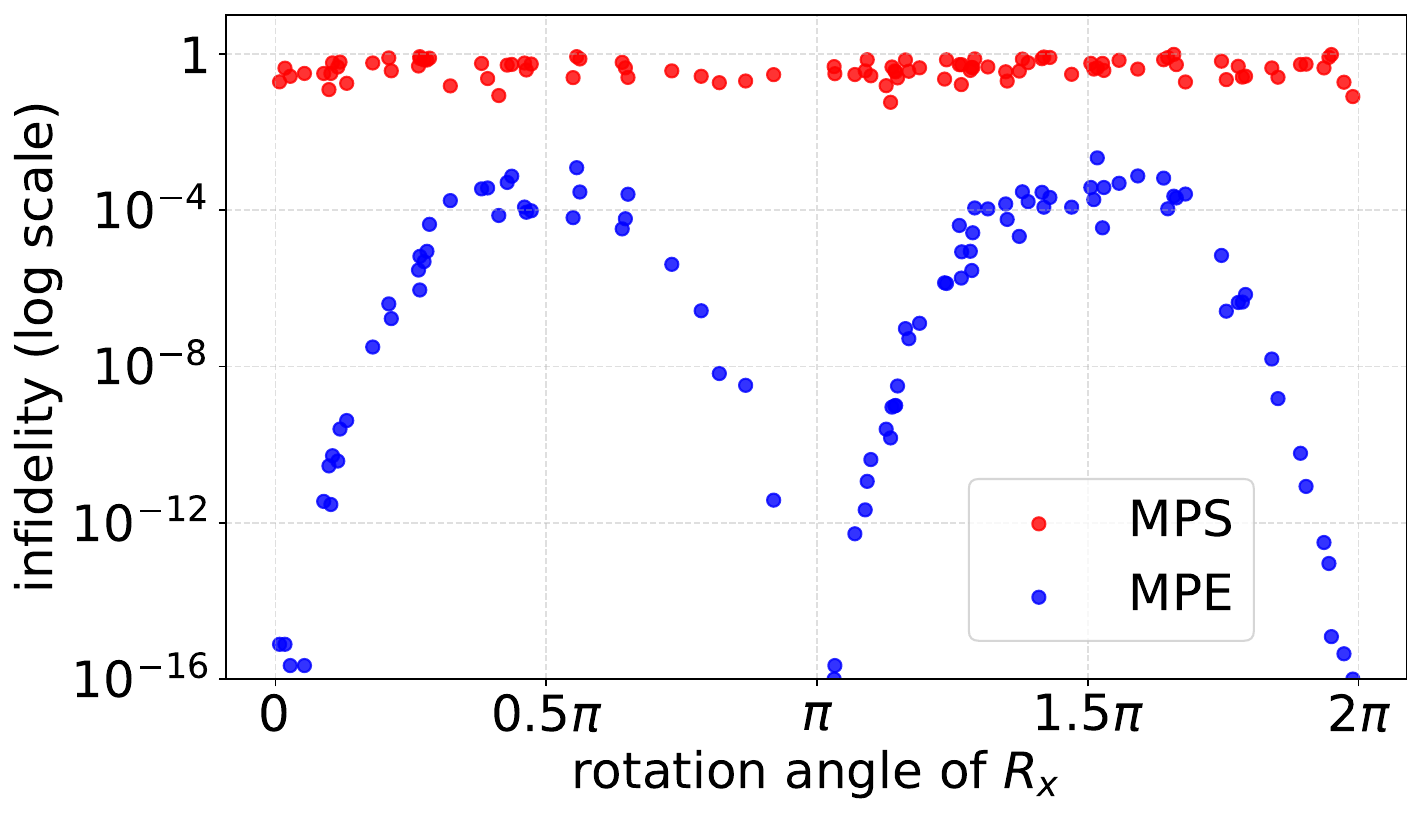}
\caption{Approximation errors of the MPS- and MPE-based methods for the $14$-qubit Ising model with maximum bond dimension $\chi=40$. 
The horizontal axis represents the rotation angle $\varphi$ of the gate $R_x(\varphi)$. 
The circuit is constructed by repeating the single Trotter step shown in Fig.~\ref{fig:ising}, and consists of $7$ steps.}
\label{fig:ising-fidelity}
\end{figure}

\begin{figure}[t]
\includegraphics[width=0.48\textwidth]{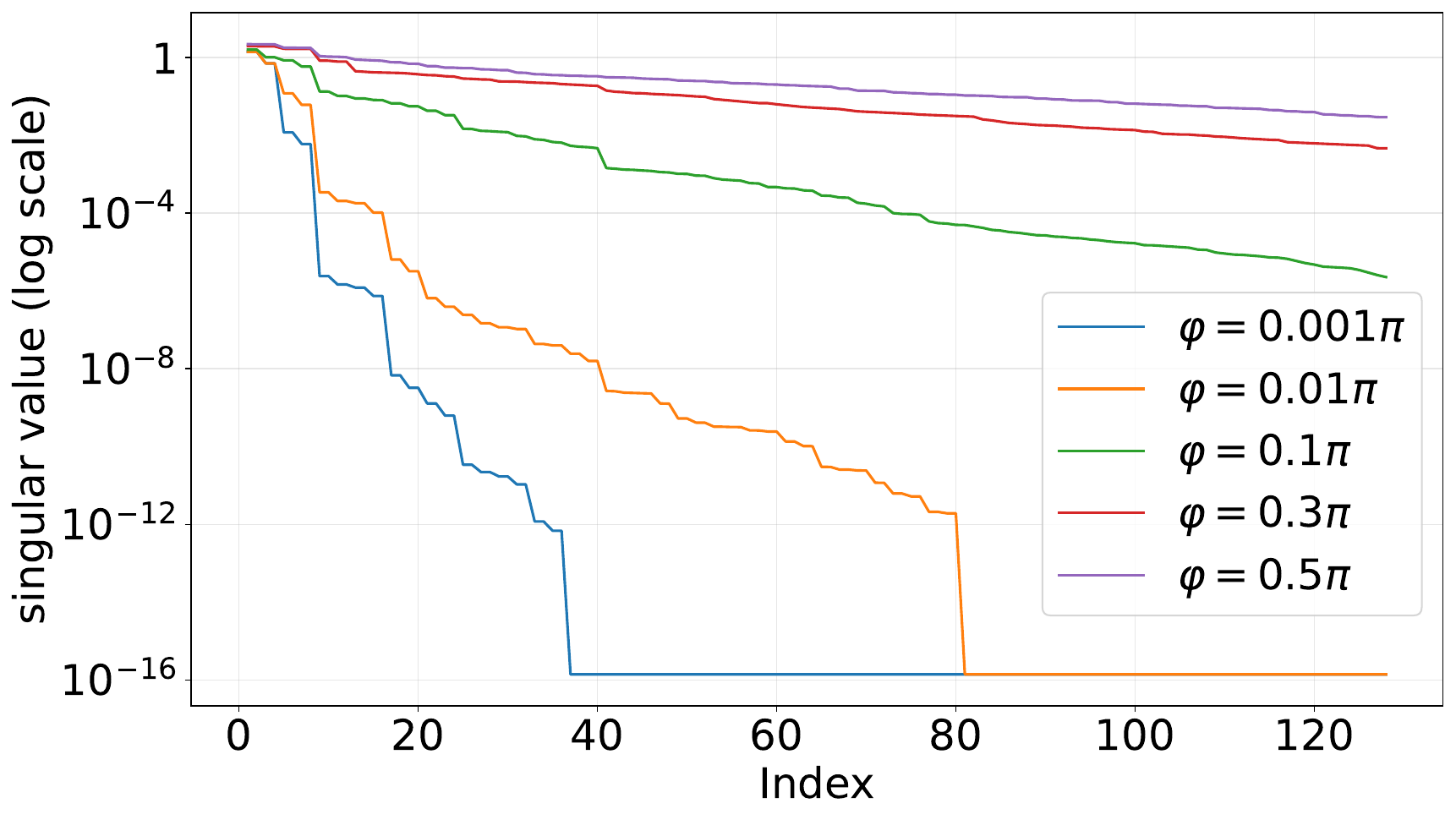}
\caption{The distribution of singular values during the MPE-based simulation for the $14$-qubit Ising model with $7$ Trotter steps. 
The horizontal axis represents the index of the singular values sorted by magnitude.}
\label{fig:ising-singular-value}
\end{figure}

\begin{figure}[t]
\includegraphics[width=0.48\textwidth]{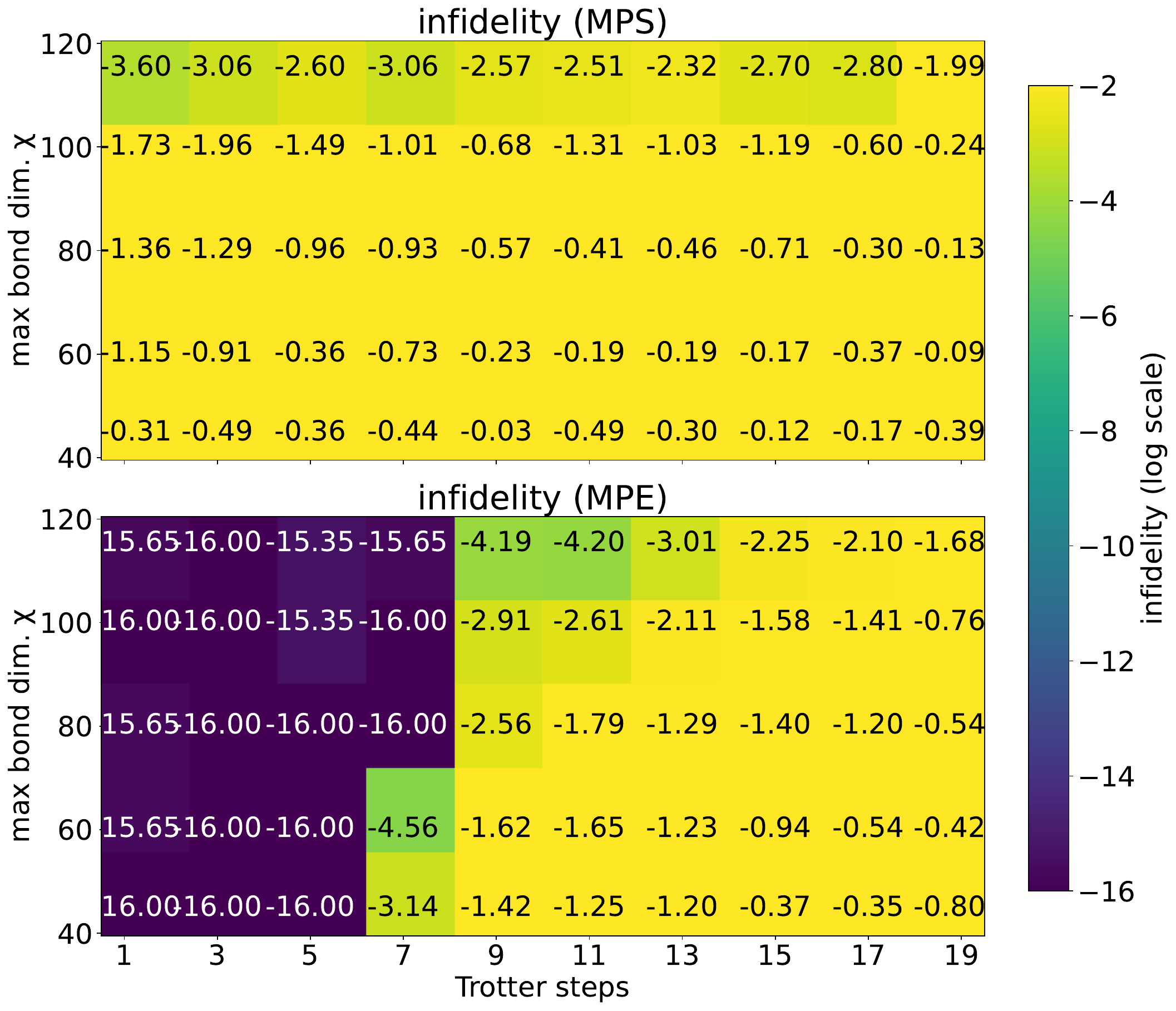}
\caption{Heat map of the infidelity (on a base-$10$ logarithmic scale) for the Ising model circuits with the rotation angle $\varphi = \pi/2$ in the gate $R_x(\varphi)$. 
Post-selection is applied to all qubits except the boundary qubits. The circuit is constructed by repeating the single Trotter step shown in Fig.~\ref{fig:ising}.}
\label{fig:ising-heatmap}
\end{figure}

We next consider quantum circuits used to simulate the time evolution of the one-dimensional Ising model. 
Figure~\ref{fig:ising} illustrates a single Trotter step that is repeated to build the full circuit used in our experiments.
Time-evolution circuits are constructed using standard decompositions in product form for local Hamiltonian dynamics \cite{Trotter1959, Suzuki1991}.
The Hamiltonian of the system is given by
\[
H = -\sum_{j} J_j Z_j Z_{j+1} - \sum_{j} h_j Z_j - g \sum_{j} X_j .
\]
In the circuit representation shown in Fig.~\ref{fig:ising}, the nearest-neighbor interaction term $J_j Z_j Z_{j+1}$ is implemented by the sequence of CNOT--$R_z(\theta_j)$--CNOT gates, while the transverse-field term $gX_j$ is represented by the single-qubit rotation $R_x(\varphi)$.
The longitudinal-field contribution $h_j Z_j$ is represented by the single-qubit rotation $R_z(\varphi_j)$.
The angle $\varphi$ is taken to be identical for all qubits, while the angles $\varphi_j$ are chosen independently for each qubit from a uniform distribution on $[0,2\pi)$.
All rotation angles $\theta_j$ are fixed once in advance and used consistently in all Trotter steps, using the parameter values provided in QASMBench\cite{Li2020}.

These circuits provide a structured setting in which correlations develop along the spatial direction as the circuit depth increases through repeated Trotter steps. In numerical experiments, the circuit used for comparison is constructed by repeating the single Trotter step shown in Fig.~\ref{fig:ising} multiple times, so that the circuit depth increases with the number of repetitions.

We compare the approximation accuracy of the MPE-based and MPS-based methods under bond-dimension truncation for the circuits with various rotation angles $\varphi$ of $R_x(\varphi)$, while keeping the independently generated values of $\varphi_j$ for $R_z(\varphi_j)$ fixed throughout the simulation. 
In this analysis, the initial state is taken to be a highly entangled MPS, as in the previous random circuit experiments, to probe a regime where the MPS-based simulation approaches its limit.

The post-selection is imposed on all qubits except for the boundary qubits, as in the case of random circuits, in order to highlight its different effects in the MPS- and MPE-based approaches, where it has little impact in the former but improves the accuracy in the latter.

The corresponding results are shown in Fig.~\ref{fig:ising-fidelity}. 
Under these conditions, the MPE-based method consistently yields lower approximation errors than the MPS-based approach because post-selection substantially reduces the truncation error in the MPE contraction. 
Furthermore, the performance of the MPE-based method is strongly dependent on the rotation angle $\varphi$ and improves significantly when $\varphi$ is close to a multiple of $\pi$, while the accuracy of the approximation deteriorates for intermediate values such as $\varphi \approx 0.5\pi$.

This dependence can be understood from the structure of the quantum circuit for the Ising model.
When the rotation angle $\varphi$ is exactly a multiple of $\pi$, the gate $R_x(\varphi)$ reduces to the identity or the Pauli-$X$ operator, so that no additional mixing between computational-basis states is generated.
In this limit, the temporal-compression procedure can successively absorb local tensor blocks. As a result, nearly the entire circuit depth is eliminated during temporal compression, leaving only a single nontrivial time slice in the compressed tensor network.
Consequently, only a few singular values remain nonzero throughout the MPE-based simulation.

When $\varphi$ is close to a multiple of $\pi$, the same tendency persists approximately, leading to a rapidly decaying singular-value spectrum. 
In contrast, for intermediate values of $\varphi$, the singular-value spectrum becomes much broader, so truncation eliminates more significant contributions.
This behavior is confirmed by the distribution of singular values in the MPE-based contraction shown in Fig.~\ref{fig:ising-singular-value}, and it explains the strong dependence of the approximation accuracy on the rotation angle $\varphi$ observed in Fig.~\ref{fig:ising-fidelity}.

Even in the case that the rotation angle $\varphi$ of $R_x(\varphi)$ is $\pi/2$, which produces the lowest accuracy among the angles considered, post-selection significantly improves the performance of the MPE-based method, as shown in Fig.~\ref{fig:ising-heatmap}. 
In this setup, the dependence on the bond dimension $\chi$ becomes relatively weak, and exact contractions remain possible over a finite range of circuit depths due to the reduction of effective degrees of freedom induced by post-selection.
Furthermore, even beyond this regime, the MPE-based method provides a more accurate approximation than the MPS-based approach under comparable truncation conditions. 
These results demonstrate that post-selection can substantially alter the effective contraction complexity in the MPE framework.

\section{Discussion}

The numerical experiments presented in this work reveal several characteristic features of the MPE formulation.

First, our results indicate that the growth of bond dimensions in MPE is strongly influenced by the circuit depth and by the degree of post-selection applied to the output state.
This contrasts with MPS-based simulations, where the dominant factor is the spatial entanglement generated during the evolution. 
Although this feature suggests that MPE may be less sensitive to the saturation of spatial bond dimensions in highly entangled initial states, the actual performance depends strongly on the circuit structure and the resulting temporal singular-value spectrum.

Second, the numerical results confirm that post-selection affects MPE and MPS in fundamentally different ways, because it constrains temporal degrees of freedom in MPE but does not directly constrain spatial bonds in MPS.
This distinction motivates further investigation of MPE for applications in which post-selection is intrinsic, including measurement-based protocols and circuits with mid-circuit measurements.

Third, the numerical behavior of MPE depends sensitively on the structure of the circuit, particularly on the singular-value spectra arising during temporal contractions. 
Circuits with gate patterns that generate flat singular-value distributions require larger bond dimensions to maintain accuracy, whereas circuits with more rapidly decaying spectra can be approximated accurately with smaller bond dimensions.
This dependence reflects a structural property of the MPE representation, rather than a universal advantage over MPS. 
In particular, circuits that generate rapid growth of temporal entanglement may exceed the practical limits of MPE-based contraction.

In conclusion, these results identify MPE as a complementary contraction strategy whose performance depends on the temporal singular-value structure of the circuit and can benefit substantially from post-selection.
This is consistent with the general view that the simulability of tensor networks is strongly dependent on the circuit structure, the entanglement distribution, and the contraction order. 
Future work includes extending the MPE framework to circuits with mid-circuit measurements, exploring hybrid contraction strategies that combine spatial and temporal representations, and developing adaptive schemes for selecting optimal contraction orders based on circuit structure.

\acknowledgments
This work is supported by Center of Innovations for Sustainable Quantum AI (JPMJPF2221) from Japan Science and Technology Agency (JST), Japan.


\bibliography{references}
\end{document}